**David Elijah Packer: cluster variables, meteors and the solar corona**

**Jeremy Shears**

**Abstract**

David Elijah Packer (1862-1936), a librarian by profession, was an enthusiastic amateur astronomer who observed from London and Birmingham. He first came to the attention of the astronomical community in 1890 when he discovered a variable star in the globular cluster M5, only the second periodic variable to be discovered in a globular cluster. He also observed meteors and nebulae, on one occasion reporting a brightening in the nucleus of the galaxy M77. However, his remarkable claims in 1896 that he had photographed the solar corona in daylight were soon shown to be flawed.

**Biographical sketch**

David Elijah Packer was born in Bermondsey, London, in 1862 April (1) the eldest child of Edward Packer (1820-1896) and Emma (Bidmead) Packer (1831-1918) (2). David attended the Free Grammar School of Saint Olave and Saint John at Southwark, where he performed well, especially in arithmetic, algebra and general mathematics (3).

Edward Packer was a basket maker, as was his father and grandfather before him, the Packer family originally coming to London from Thanet, in Kent. The English basket-making industry was in decline during the second half of the nineteenth century due to the availability of cheaper imports from the continent. In spite of this, Edward was still in business in 1881, but the prospects for his son continuing in a trade that had made a living for generations of his ancestors were probably limited (4). Thus at the age of 18 we find David Elijah Packer working as an Oilman's Assistant, supplying fuel for the oil lamps of London (5). However, ten years later, in 1891, and still living with his parents, he had apparently returned to the world of education as a "Student of Natural Science and Mathematics" (6). At this time it was most unusual to leave paid employment to return to study. It was even more unusual to do so at the age of nearly thirty, perhaps suggesting his academic talents had been recognised, but certainly indicating an enthusiasm for science. In 1892 he worked at Cambridge, although there are no details of his activities there (7). He later commented to family members: "professional jealousy soon drove me from Cambridge and I became an exile" (8).

Packer soon found alternative employment, as a librarian at Birmingham Central Free Library. He married Sarah Ann Day (1861-1921) on 1893 Dec 10 at Southwark, London (9), and the couple went on to have seven children (Figure 2) (10). The marriage took place around the same time as his move to Birmingham. In fact his letter to the BAA *Journal* written just four days after the wedding gave the Library's address (11). He later became Librarian at the city's Stirchley & Bournville Free Public Library (Figure 3) (12). During the First World War he assumed oversight of three libraries due to staff shortages caused by men serving at the Front. He was responsible for organising the first "open access" public library in Birmingham in which books could be viewed on shelves, rather than presenting a librarian with a list of titles that they would then fetch from the book room (8).

Packer lived in south Birmingham for the rest of his life. The newly married couple initially set up home at 33 Croydon Road, Bournbrook, then in 1903 they moved to a new home a few streets away at 71 Oak Tree Lane, Selly Oak. With an expanding family they rented a larger property a short distance away at 2 Lansdowne Villas, 864 Bristol Road. When Packer retired from library service he was allocated a new council house in Northfield, just south of Selly Oak, where he lived with his daughter Ruth, his wife already having died. When Ruth married, he moved into her new home at Sunningdale, Ashmead Drive, Rednal (now on the south-western edge of Birmingham). This was adjacent to the Lickey Hills, where he enjoyed walking until his death in 1936 (8).

Quite when Packer became interested in astronomy is not known. He was certainly observing the night skies in 1882, at the age of twenty, and six years later, in 1888 June, he was using a 4½-inch (11.4 cm) refractor to follow Sawerthal's Comet (C/1888 D1) from Bermondsey (13). The instrument was of the dialyte design, which was popular at the time as it was cheaper to build than conventional achromatic telescopes. It comprised a large crown glass objective with a much smaller flint glass downstream to make an achromatic lens, flint glass being more expensive. Dialtye refractors were advocated by Packer's friend and mentor, Herbert Ingall FRAS (1846-1903), who lived not too far away in Champion Hill, south London. Packer visited regularly Ingall to join him for observing sessions (14). Ingall had developed a reputation as a careful lunar observer from the 1860's onwards, at one time collaborating with another renowned selenographer, W.R. Birt (1804-1881). Ingall was an original member of the BAA, serving first on its provisional committee and then on its first Council in 1890 (15).

Packer became a member of the BAA in 1892 November. Never a Fellow of the Royal Astronomical Society, he was however at one stage a doorkeeper at the Society's rooms in Burlington House (16). He was keen to popularise "astronomy for the schools and working classes" (17); for example, during the apparition of 1P/Halley he was often seen on Weoley Common in Birmingham offering views of the comet to the public (8).

**Messier 5 and the cluster variables**

During the spring and summer of 1890, and still living in London, Packer noticed that a 10[th] magnitude star near the edge of M5, the globular cluster in Serpens, appeared to be variable in brightness. His initial observations were made with his 4½-inch refractor and subsequently corroborated using Ingall's 10-inch instrument. Finding no reference to a known variable star in the astronomy books to which he had access, he contacted Prof. Ralph Copeland (1837-1905) at the Royal Observatory Edinburgh and T.H.E.C. Espin (1858-1934) at Tow Law, County Durham. Espin soon confirmed its variability. By coincidence, whilst attending a meeting of the RAS Packer saw exhibited two photographs of M5 taken by A.A. Common (1841-1903) in 1890 April and May, each of which showed the star at a different brightness (18). Packer announced his discovery of the new variable in the *English Mechanic* of June 27 (19) and in the *Sidereal Messenger* (20). His observations with Ingall's telescope also revealed a second star in M5 that he suspected was variable and his sketch of the cluster showing both variables is shown in Figure 4. Once again, reference to Common's photographs confirmed this to be the case. Noting his suspicion that other stars in the vicinity of M5 might also be variable, Packer suggested that others might like to take up the challenge of examining this cluster and others for variable stars (21).

Packer's reports of the two new variables were noticed by Williamina Fleming (1857-1911) at Harvard College Observatory who used the Observatory's plate collection to confirm their variability, which Professor E.C. Pickering (1846-1919) announced in *Astronomische Nachrichten* (22). Pickering was already familiar with Packer's work as the two had previously corresponded about variable stars, prompting Pickering to send him a manuscript of his latest variable star catalogue (23). The Harvard confirmation, citing Packer's observations must have come as a relief to him as he was becoming concerned that Common was being credited with the discovery, not because of any claims made by Common himself, merely that Packer's announcements, made by a relatively unknown individual, had been overlooked by many (24). Even a few years later, it was still not universally known that Packer had made the discovery. For example when E.E. Barnard (1857-1923) wrote a paper on the variables in M5 in 1898, he added the following *Postcript*: "Since writing my paper on the variable stars of the cluster M5, my attention has been called…to the fact that two of these variable were originally discovered by Mr. D. Packer, of London, in 1890……I am glad to be able to make this correction for Mr. Packer certainly deserved credit for this discovery" (25).

Packer's role in discovering the first variable in M5 was also highlighted by Solon Bailey (1854-1931) of Harvard College Observatory in his definitive 1917 review of *Variable Stars in the Cluster Messier 5* (26). Helen Sawyer Hogg (1905-1993), noted for her pioneering research into globular clusters and variable stars, also credited Packer (27). Packer's discovery was particularly important as it was only the second periodic variable to be found in a globular cluster, the first one being discovered in M3 in 1889 by W.H. Pickering's (1858-1938) expedition to southern California from Harvard College Observatory (28).

More than 100 stars in M5 are now known to be variable in brightness, the vast majority of them belonging to the RR Lyrae class. RR Lyrae stars, sometimes referred to as "Cluster Variables", are pulsating stars similar to Cepheids and as such can be used as standard candles to measure distances since the relationship between their luminosities and periods is well known. They have periods typically in the region 0.2 to 2 days. For many years Packer's two variables were classified RR Lyrae stars, the first of the class to be discovered in a globular. However, more recently another class of pulsating variable has been recognised, the W Virginis variables (29), also known as a Type II Cepheids, with a period of 1 to 35 days to which both of Packer's stars have been assigned. The two stars have been studied extensively over the ensuing years and are known as M5 V42 (30), with a period of 25.738 days, and M5 V84, with a period of 26.42 days (31) (Figures 5 and 6).

Early in the year following Packer's discovery, 1891, he announced the detection of two more suspected variables, this time near the open clusters M103 and NGC 475 in Cassiopeia, although no confirmation of either was forthcoming (32). In spite of his interest in discovering variable stars, there is no evidence that he pursued the systematic observation of known variables, other than variations in Betelgeuse which will be discussed next (33).

**The variability of Betelgeuse**

The variability of Betelgeuse was first noticed by Sir John Herschel in 1836 and described in his *Outlines of Astronomy*, published in 1849. Packer's announcement in the *English Mechanic* in late 1902 of an unusually bright state of Betelgeuse triggered a flurry of activity amongst observers. "On the night of Oct. 15", he wrote, "I was astonished to find that the bright star Betelgeux ($\alpha$ Orionis) had greatly increased in magnitude, being superior to Capella, and only slightly inferior to Sirius in point of brilliancy" (34). Copeland of Edinburgh issued a circular to observers around the world notifying them of Packer's observation. The respected variable star observer J.E. Gore (1845-1910), who had directed the BAA Variable Star Section (VSS) from its inception in 1890 until 1899, provided confirmation, finding it to be about as bright as Capella on October 16 (35). Following Packer's announcement many others confirmed that it appeared to be bright to them too, although quite how bright varied from one observer to another. For example, Ernst Hartwig (1851-1923) of the Bamberg Observatory also found it slightly fainter than Capella (36). By contrast, W.T. Lynn (1835-1911) observing from Blackheath, London, on October 19 said "It did not seem to me to be brighter than usual…very much less bright than Capella" (37).

A few months later E.E. Markwick (1853-1925), the incumbent VSS director, analysed all the available data and concluded that indeed Betelgeuse was brighter than in previous years, perhaps by half a magnitude, and probably brighter than it had been since 1888. However, there was considerable scatter in the data, especially early in the season (October-November) when the star would have been low in the sky compared to its comparisons (38). Figure 7 shows Markwick's light curve, largely based on data from VSS observers. He pointed out that data from three observers appeared to be discrepant: Packer, R.T.A. Innes (1861-1933) (39) and Mary Orr (1867-1949) (40). Neither Innes nor Packer had provided details of the sequence used, whereas Orr, uniquely, had used Sirius as a comparison which Markwick thought "may account for her results".

This episode illustrates some important points about variable star observation. Observers need to use the same comparison stars and sequences. Fortunately most VSS observers had used a common sequence, but some reports did not specify the sequence or the comparisons used. As Markwick noted, estimating Betelgeuse is particularly challenging because there are few comparison stars available and these are widely separated, which can also introduce differential atmospheric effects when the stars are at different altitudes. Furthermore the star's red colour means that its brightness can be perceived differently by the eyes of different observers. Referring specifically to the events surrounding Betelgeuse in 1902, Joel Stebbins (1878-1966), who went on to perform pioneering photometry on the star with the newly developed selenium cell, commented that "it is an interesting psychological fact that while a sudden brightening of such a star will cause scores of astronomers and laymen to watch it carefully, an equally conspicuous decrease in light occasions little or no comment" (41). The confirmation by several observers of Packer's announcement of an exceptionally bright state might illustrate another potential pitfall in variable star work: observer bias.

Another aspect of Packer's observations of Betelgeuse in late 1902 and 1903 was more controversial. On several occasions he reported rapid variations in brightness, or scintillation, of a few tenths of a magnitude over a few seconds or minutes (Figure 8). He also recorded changes in the degree of "ruddiness" of the star. Initially he put these effects down to the atmospheric effects, but his studies suggested to him that they were more pronounced than in Aldebaran and Procyon, which, he thought, "shows that atmospheric effects are only partly responsible for these changes, and for the rest we must look to the peculiar constitution of the star's light, which is now exalted to threefold brilliancy" (42). A paper on the subject was read at the BAA meeting of 1903 February 25 on his behalf by the Association's Secretary, A.C.D. Crommelin (1865-1939), who expressed scepticism, suggesting it was likely due to atmospheric scintillation. Moreover, although the actual size of Betelgeuse was not then known, it was understood to be a "giant" star, larger than the Sun, which led Crommelin to comment of such a giant star that "it was a little incredible that its light could sensibly vary in a few minutes" (43).

We now know that Betelgeuse does not exhibit these rapid brightness variations and that what Packer saw was probably atmospheric in origin. He continued to monitor the star over the years and to report any unusually bright episodes. He also suggested that spectroscopic observations of the star might shed light on its behaviour. He himself used a small direct vision spectroscope in combination with his telescope (44) and took a great interest in contemporary developments in spectroscopy.

**The Espin-Peek Phenomenon**

Packer studied the reports of variable stars observations made by Espin, at Tow Law, and Sir Cuthbert Peek (1855-1901) at his Rousden Observatory, near Lyme Regis. He noticed that many of the long period variable stars they observed bore resemblance to novae in two respects: (1) by the presence of emissions lines in their spectra, particularly hydrogen and helium lines, and (2) in displaying a nebulous halo most evident when they were near minima. In 1902 Packer wrote at length on his investigation correlating Espin's observations of the emission lines with Peek's observations of nebulosity. He coined this 'The Espin-Peek Phenomenon' (45). Later he invoked the presence of emission lines in red variable stars recorded by Fr. Angelo Secchi (1818-1878) at Rome in 1868 and 1869 in support of his argument and added his own observations of emission spectra (46). It appears that nobody followed up on Packer's ideas, but they were likely flawed since long period variables are not associated with nebulosity at minimum, or at any other time – the reported nebulosity might have been an optical aberration. On the other hand, when a nova fades, the gas blown off by the explosion can give rise to a nebulous shell surrounding the star which presents emission lines. At about the time of Packer's publication of the Espin-Peek Phenomenon a nebula had been detected surrounding Nova Persei 1901 (GK Per). Similarly, Packer would also have been well aware of the emission line nebula observed following the outburst of Nova Aurigae 1891 (T Aur), which appeared as a small disc surrounding the star.

The cause of a nova outburst was hotly debated in the astronomical community at the time, with some suggesting it resulted from two stars colliding. Packer took a different view. Stimulated by his thinking on the appearance of nebulae around novae, he proposed that it was the nebula passing in front of the star that caused the star's light to brighten: "Just as a mirage is a remote object enormously magnified by altered atmospheric conditions, so a temporary star is a spurious and greatly magnified image of some remote star, and due to abnormal conditions between our Solar System and the star. The effect is an optical one" (47). Such an idea, although incorrect as an explanation of the nova phenomenon, is curiously reminiscent of the modern understanding of gravitational microlensing events which can cause stars to brighten by several magnitudes over a period of a few days (48).

**Variable nebulae and a possible flare in M77**

Packer's interests were not confined to *stars* which vary in brightness, for soon after his discovery of the cluster variables near M5, he penned a note about possible variations in brightness of nebulae detected with his 4½-inch refractor. His observations of M77 (NGC 1068) are noteworthy: (49).

"N.G.C. 1068 = M 77 Ceti. R.A. 2h. 35m. 31s., N.P.D. 90° 36.7'. Webb calls this faint. I found it, 1890, Sept. 20, v. bright, with a brilliant nucleus equal to a star 8 mag. This subsequently faded, till on Oct. 15 the nucleus was scarcely perceptible".

It is impossible to know the detection limit of Packer's 4½-inch refractor, which he typically used at a magnification of x40, under the skies of south London, which he admitted elsewhere were not ideal. A popular online limiting magnitude (LM) calculator (50) suggests an LM of about 12.5 for a modern refractor of this size, with a naked eye LM of 5.5. Now, Packer's telescope had no optical coatings which might have reduced the LM further to, say, 12.0. Such would be the LM for a stellar object on a dark background, but of course in the case of M77 the background would have been the galaxy itself, which might further reduce the detection limit to 11.5. Thus it is possible that the amplitude of the event observed by Packer in M77 was about 3.5 magnitudes (or more, considering it was fading from the time of his first observation). Gerard de Vaucouleurs (1918-1995) (51) estimated the brightness of the nuclear region of M77 to be about magnitude 12.0, setting the detection limit beyond which the object would have emerged with the background, which led him to suggest the amplitude might have been about 4 magnitudes.

Thus, putting all this together, we have a star-like object which might have declined 3.5 to 4 magnitudes in 25 days, or about 0.14 to 0.16 magnitudes per day. M77 is now known to be a Seyfert galaxy which led de Vaucouleurs to raise the intriguing question as to whether Packer might have observed the fade following a flare-up of the nucleus (51). De Vaucouleurs further speculated that the flare, if real, might have been caused by the accretion of a star by a central black hole. However, by contrast to many other Active Galactic Nuclei, M77, as a Type 2 Seyfert galaxy, is not known for large visual variations; indeed if there are variations at all, they are much less than 1 magnitude (52).

An alternative explanation is that Packer observed a supernova. De Vaucouleurs also considered this hypothesis, but thought it was unlikely due to the rapid decline (Type 1a supernovae, which show the fastest decline, have decline rates of 0.05 to 0.085 magnitudes per day in the V-band over the first 20 days (53)). No supernovae have been observed in M77 since records began in 1885 (54).

It should be noted that Packer also suspected that both M31 and M63 (55) have variable nuclei (56). The case of the brightening of the nucleus of M31 is particularly intriguing: "on the night of [1890] Sept. 15 I also noted a singular phenomenon in the nucleus itself, which appeared unusually brilliant and shining with a star-like lustre" (49). Assembling all of his observations, he even believed he had evidence for a periodic variation of 45.2 days.

Packer was certainly not the first person to describe variability in the nucleus of M31: there had been reports throughout the nineteenth century by many reliable observers, although others disagreed. One long-time proponent was Ernst Hartwig who still claimed variability in 1888 (57). Isaac Roberts (1829-1904) announced that he had secured definitive photographic evidence for variability in 1891(58) and it was Roberts's revelations that stimulated Packer to come forward with his own observations. Roberts's reports were confirmed by W. Seraphimoff of the Pulkova Observatory at St. Petersburg (59). However, the question of M31's variability was finally resolved by E.E. Barnard in 1898, whose careful photographic measurements disproved the idea (60).

One explanation as to why many people thought that the nucleus of M31 was variable is that the view through the eyepiece is highly dependent on the observing conditions (this also applies to photographic observations, with the additional variables of the exposure time and photographic process). Under some conditions of seeing and transparency, the nucleus can appear very compact and bright – almost star-like, to use Packer's term - whereas at other times it can appear more extended and diffuse. Such is the case for other diffuse objects including galaxies, nebulae and comets.

These considerations do, therefore, raise a question mark over the reliability of Packer's observation of the M77 event. Nevertheless, there remains a tantalising possibility that an amateur astronomer living in south London may have been the sole witness of an important cosmological event.

Packer maintained a broad interest in what today would be called "Deep Sky Objects". In 1903 he wrote to the BAA Council proposing the Association set up a section for the "study of nebulae and clusters", pointing out that sections existed for almost every other aspect of observational astronomy (61). He even put himself forward as director and set out a programme of work for the section, where some his personal interests were evident:

> (1) Study of the internal and external structure of nebulas and star-clusters and their comparison for the discovery of new types.

> (2) Study of the alleged variability of the nuclei of nebulas and of entire nebulas.

> (3) Study of the variability of the more conspicuous stars in the clusters for detecting possible relations between their range and periods.

> (4) Search for new stars in and around the region of star-clusters and nebulas.

> (5) An examination of the great swarms of nebulas in Leo and Virgo by means of Herschel's methods of sweeps

> (6) Search for those spurious nebulas or "glows" seen in the neighbourhood of new stars and long period variables when at minimum during abnormal atmospheric conditions (62).

> (7) Drawings and delineations of our greatest nebula and cluster—the Milky Way.

Council did not take up Packer's proposal and it wasn't until nearly 80 years later, in 1981, that the BAA finally established a Deep Sky Section under the directorship of Ron Arbour (63).

**Nebulous meteors, earthquakes and temporary stars**

Packer was an enthusiastic observer of meteors, contributing to the work of the BAA Meteor Section, as well as sending in reports of observations of meteor showers and fireballs to the *English Mechanic*. Increasingly, his reports were of unusual meteoric activity, such as sonic booms, "nebulous meteors" and obscurations. For example, we read: (64)

"I should like to call attention to a rather uncommon type of meteor which generally appears in the form of a nebulous cloud, at, presumably, the radiant point of some shower"

Thus on 1888 Oct 13 he reported: "Brilliantly clear, meteors abundant, a small, round, stationary nebulous meteor of a greenish tinge (precisely like a planetary nebula in appearance) appeared"

Packer took it upon himself to catalogue the appearance of nebulous meteors and encouraged others to look out for them. In undertaking observations of this type, he advocated proper dark adaptation and "in observing I recommend all abstinence from narcotics or alcoholic stimulants at all times. I never use them, and find the eyesight greatly improved and strengthened thereby" (65). Surely sound advice for all observers even today!

Whilst some were sceptical about Packer's nebulous meteors (66), others, including well-known observers such as Julien Péridier (1882-1967), also recorded examples of the phenomenon from time to time. However, W.F. Denning (1848-1931) of Bristol, one of England's most respected meteor observers, sought to distance himself from Packer's observations, especially when Packer cited one of his own observations. Denning wrote: "These supposed bodies are really not worth any discussion; but I feel bound to protest now that Mr. Packer….quotes a meteoric observation of mine in support of his views…. I can assure your readers that the meteor referred to by Mr. Packer as seen by me on July 24, 1900, was a different thing altogether to the very extraordinary appearances discovered by him, and I really do not think there is any prospect of my corroborating any of his results as long as I can keep my imagination under decent control" (67).

So what should we make of all this? Packer is by no means the only person to have described nebulous meteors, and there were others who reported them too, but the sheer number he reported clearly caused disbelief amongst many. Even today meteors are occasionally reported by reliable observers that do not have the sharply defined contours of most meteors and are somewhat fuzzy in appearance. They might be highly friable objects that undergo several stages of disintegration (68). Perhaps the matter will be resolved by the increasing number of video detection system being deployed.

Packer's credibility as a reliable observer of meteoric events might have been irrevocably tarnished in Denning's mind by his numerous reports of nebulous meteors, for when Packer reported an audible detonation following the fireball of 1905 Jan 27 (the second such event he reported that month (69)), Denning was highly sceptical. Observing from Selly Oak, Packer found that "its light was so vivid that it illuminated the sky with the brilliancy of the full moon….. At 1m. 30s. after disappearance a wave of sound like distant rolling thunder was heard; this lasted for six or eight seconds, then gradually subsided for twelve seconds, when a second wave of sound of even greater intensity came up, finally subsiding into silence by two minutes" (70). Others in southern England had seen the fireball and Denning pieced together its likely trajectory, pointing out that the trajectory Packer had given was not consistent with the other observations. Moreover, Denning's calculations suggested that if a sonic event had been associated with the fireball it would have occurred some 12 minutes after the visual event (71).

In addition to Packer's animated reports of nebulous meteors and audible detonations, his descriptions of the major meteor showers became increasingly colourful. A particularly ebullient description of the 1904 Perseids prompted one person to comment testily: "Is it not possible for Mr. Packer to watch an ordinary meteoric display without transforming it into a magnificent exhibition with most unusual accompaniments, such as flashings, coruscations, nebulous meteors, &c, &c.?" (72)

If Packer's observations of unusual meteoric activity were not met with enthusiasm by everyone, his ideas on the links between meteors and other natural phenomena raised more than a few eyebrows. Under the rather grand title of "*A New Cosmical Law*", he set out his theory that there might be a correspondence between the location of meteor radiants in the sky and the position of novae. He went on boldly to conclude that "it may now be accepted as a provisional law that all new or temporary stars represent centres of meteoric radiation" (73). He announced what he believed to be several new meteor shower radiants which were associated with old novae, naming the supposed shower after the nova. Thus we had the Nova Perseids, associated with N Per 1901, the Nova Cassiopeids, associated with Tycho's star of 1572 (which is now known to be not a nova, but a supernova), the Nova Aurigids (N Aur 1891) and the Nova Cygnids (N Cyg 1876) (74).

Many had strong objections to this "new law", notably W.H.S. Monck (1839-1915), as lacking any semblance of credibility (75). Packer's clarification of his position, in response to Monck's criticism, did not mollify matters: "I have not for a moment supposed that these meteors come straight from the new stars referred to, but simply stated the fact — that these two celestial phenomenon are constantly found to coincide in time and place" (76).

Not content with linking meteors and novae, Packer went on to suggest an association between the appearance of meteors and the occurrence of earthquakes. Thus observing from Bournbrook in Birmingham, "on the night of the recent earthquake in Scotland (Sept. 17 [1891]) several remarkable meteoric appearances were seen here an hour before the shock" (77). He cited several other examples of earlier earthquakes that had occurred at about the same time as unusual meteoric activity, concluding "that a relation between earthquakes [and meteors] exists, there can be no doubt, although the law governing the relation remains to be discovered".

**Comets and solar activity**

Packer was a keen observer of any comet that happened to be undergoing a bright apparition. However, as with his meteor work, he opened himself up to criticism when he moved from merely reporting what he saw to more theoretical considerations. Thus he announced in a letter to the BAA *Journal* in 1896, entitled *Comets and Sunspot Maxima,* that he had found a correlation between the appearance of comets and the solar cycle (78). In the letter he gave some examples of bright comets of the past that had been observed close to sunspot maximum. This was not merely anecdotal for he had tried to cross-correlate comets recorded in various catalogues with solar activity, by plotting the events on graphs. As well as finding groupings of comets which occurred in an 11-year period, he believed there was evidence for a secondary period of around 60 years. He did not present the data in the letter, but he ended with a flourish and a hint of more to come: "That a mutual relation between apparitions of comets and sun-spot maxima exists is inferred and abundantly confirmed by the evidence we possess, evidence I hope very shortly to place before the Members of the Association, together with the results which I have obtained therefrom".

Packer did go on to submit "a very lengthy paper" on *A Remarkable Relation between Comets and the Solar Variation* for the BAA President, E.W. Maunder (1851-1926), announced at the 1895 June BAA meeting that such a paper been received (79). Packer was unable to attend the meeting. The paper was not read: Maunder only gave a brief summary of its contents and conclusions to the audience. Of course, Maunder was an expert in solar activity, having studied the variation in the latitude of sunspots during the solar cycle which resulted in the famous "butterfly diagram", and would have known immediately that there were problems with Packer's analysis. He was very polite in pointing out that he "could only give his own impression of his paper, and from the hasty glance which he had had of it, he did not feel at all convinced of the soundness of Mr. Packer's conclusions". Maunder's comments from the same BAA meeting as reported in MNRAS were pithier: Packer "had failed to establish his theory. The data seemed too vague and incomplete to be of much value" (80). In view of these comments it is perhaps not surprising that the paper never did appear in print in the *Journal*.

**The New Astro-Photography and the solar corona**

The 1896 March edition of the American astronomy magazine, *Popular Astronomy* (81), contained an article with an arresting headline:

NEW PHOTOGRAPHIC DISCOVERY - THE SOLAR CORONA PHOTOGRAPHED IN DAYLIGHT (82)

The author was none other than D.E. Packer, who claimed to have succeeded in doing what many astronomers had unsuccessfully striven to do for decades: photograph the solar corona out of eclipse. If true, this would be a truly sensational breakthrough and would ensure that Packer's name would be remembered in perpetuity. Hitherto, the only way to study the corona was during a total eclipse and it was common for observatories, organisations and individuals to organise expeditions often halfway around the world, often under challenging conditions, to benefit from a few minutes of totality.

So how had Packer apparently succeeded where so many before, including William Huggins (1824-1910) and George Ellery Hale (1868-1938), had failed? Packer was a keen photographer and had set up a darkroom shortly after moving to Selly Oak. The apparatus he used for his corona photography was simplicity itself: a camera with a 4-inch (10 cm) lens – and even that he found he was later able to dispense with, to be replaced by a pinhole – and a photographic plate covered in a metal foil. The latter was the key: he employed foil made of tin, copper or lead which are "relatively transparent to solar radiation of 'high refrangibility' ". This term would have struck a chord with readers due to the excitement in the world of physics following the announcement of the discovery of the "highly refrangible" X-rays by Wilhelm Röntgen (1845-1923) the previous year.

More details of Packer's discovery were disclosed in an article in the photographic magazine, *The Photogram*, entitled *The New-Astrophotography* (83). He described how he had first entertained the idea during the spring of 1895 "that the electrical radiance from the sun and stars…should register itself upon a photographic plate, provided that the plate was excited during exposure, by the passage of an electric current". In such a way he believed he had detected a secondary spectrum from the sun comprising rays of shorter wavelength than visible light (84). Initially he applied the metal foil to the plate and passed an electrical current through it, but he later dispensed with the electrical current, believing the foil contributed sufficient electrical field itself. In 1895 August he began experimenting with capturing images of the sun itself. He interpreted the resulting impressions on the photographic plate as representing the solar corona. Examples of two such images can be seen in Figure 9, which he said show a "series of radiations, more or less prominent, emanating from a central nucleus, which was considerably smaller than the normal projected image of the solar disc, which, if shown at all, is very feebly represented, in the photograph. Very striking are the fine needle-like rays on opposite sides, nearly, but not quite, parallel to the sun's polar axis, and the great equatorial extensions, especially over the east limb" (85).

Packer was confident that his discovery was a pivotal moment in astronomy: "A new and most important field of astro-physical research has been opened up which promises great results in the immediate future. We are now able to study this new radiance in the sun and stars, to measure its relative intensity, and note its variations. Many new celestial bodies and centres of radiance may be detected thereby, which ordinary telescopic and photographic means would never have succeeded in tracing" (85).

Readers of the *English Mechanic* were privileged since Packer made his first announcement about photographing the corona in the 1896 January 24 edition (86), well before the *Popular Astronomy* article appeared. The immediate reaction of readers was one of great excitement. The Reverend Daniel Higham Sparling, FRAS, Rector of Biddulph Moor, near Congleton, said Packer's announcement "seems to me to be of such extraordinary importance, that it cannot but be that fuller information would be most welcome to the readers of '*Ours*'" (87) ('*Ours*' being the term of affection with which many *English Mechanic* readers used to describe 'their' newspaper). BAA member Alfred Buss, from Manchester, was similarly delighted: "I cannot help expressing my great pleasure at seeing another important step in advance of our knowledge of solar physics", noting that "there is a strange resemblance of the process hit on by Mr. Packer and Prof. Röntgen's method" (88).

However, it soon became clear that there was much scepticism surrounding Packer's corona photography. He submitted a paper to the BAA which was read at the meeting of the West of Scotland Branch in Glasgow on February 12, but when a number of images were shown, "suspicions…were entertained regarding the reality of the discovery" (89). A commentator in the *Journal* observed that "the news seems far too good to be true" (90) and, in a subsequent *Journal* it was concluded that "the results are certainly interesting, but it would be unwise to regard the appearances on the photographs as truly coronal, until the method has been tested during the progress of a solar eclipse" (91).

As with Packer's earlier paper on comets and the solar cycle, his paper on coronal photography was never published in the *Journal*. Furthermore, it was rejected by George Ellery Hale for publication in his *Astrophysical Journal* and no other astronomical journal, besides *Popular Astronomy*, picked it up either. The real blow was that as the weeks and months passed by nobody came forward with authoritative confirmatory experimental data (92). When William Huggins heard about Packer's results he confided to his friend Hale: "From what I know of Packer, and from what I hear, the probability seems to be that his corona is a mare's nest. Yesterday at the Royal Society Prof. Lapworth of Birmingham [Charles Lapworth FRS (1842-1920), geologist] told me it was 'all bosh' " (93). Nevertheless Huggins ordered some thin aluminium with the aim of carrying out his own tests, which he did without success. He was dismissive of the organs that had published Packer's results, saying that "the *English Mechanic* swallows camels easily" and that "Packer seems to have 'gepackt' the Editor of *Popular Astronomy*" (94).

So what do we make of Packer's results? It was realised at the time that his idea that they might be due to X-rays emitted from the sun, passing through the metal foil, could not possibly be correct. As Huggins and others pointed out, the solar radiation at the earth contains no X-rays. Instead, Huggins suggested that Packer's "coronae are in my opinion produced by the diffraction images of minute holes nearly always present in foil" (95).

Packer continued to believe that he had secured images of the solar corona in "hundreds" of metal-foil covered plates (96). In a letter in 1903, clearly frustrated that his idea had not been taken up, he summarised his research on the subject conducted between 1893 and 1895, maintaining that "I found that when the sun was 'spotted' pretty generally over his surface that a typical corona could be obtained; but when the spot groups were all on one limb, as is not infrequently the case, the coronal rays corresponded" (97).

Others after Packer continued the quest to observe the solar corona. In 1919 H.P. Hollis (1858-1939) summarised the situation: "Although attempts have been made to observe, or rather, to photograph the solar corona when the sun is not eclipsed, but without success. The difficulty arises because the diffused light of our atmosphere is stronger than that of the corona" (98). It was not until the 1930's that Bernard Lyot (1897-1952) finally achieved success using his coronagraph.

## The *Astro-Physical Station, South Birmingham*

There is a further curiosity in connexion with Packer's work on the solar corona. In his correspondence to the *English Mechanic* at the time, as well as in his article in *Popular Astronomy*, he gave his affiliation as the *Astro-Physical Station, South Birmingham*. I can find no evidence of such an institution actually having existed and there appears to be no reference elsewhere to other work it might have conducted. One wonders whether Packer might have invoked the affiliation to lend credibility to his work as it, in his mind, would unfold within the pages of the world's preeminent astronomical journals. The address of the *Astro-Physical Station* is given as North Pershore Road, Birmingham. This is not far from Packer's residences; and the Stirchley & Bournville Free Public Library, where he worked for a time, is just off the Pershore Road.

Another period when Packer invoked the *Astro-Physical Station, South Birmingham* affiliation was in 1905 July and August in a series of letters to the *English Mechanic* outlining his increasingly odd ideas about "nebulous clouds". In the first of these he discussed observations of the zodiacal light, coming to the conclusion that, contrary to the accepted idea that it was caused by sunlight scattered from small particles, it was "a form of radiation in the neighbourhood of the earth's atmosphere", which he termed the "earth's crêpe ring"(99). Moreover, he suggested that the lack of detail visible on the surface at full moon was evidence for such a ring since the thickest part of the ring, which sometimes appeared as the gegenschein, acted as an obscuring veil (100).

Packer also went on to describe what he thought were "Great Nebulous Clouds" in Leo (101) and Scutum (102). In the case of the former, a "feebly luminous area, with Leo as a centre, is found to extend between 40° and 50° on every side, but brightest in the region of the Ecliptic". He proposed four possible explanations for this: it could be an extension of the milky way, or perhaps the combined glow of innumerable faint nebulae (many of which were known to exist in the area between Leo and Virgo, which we now know to be galaxies), or a "continuous magnetic radiance from the great Leonid meteor radiant", or, lastly, an "exaltation of the earth's ultra-atmospheric ring" (although he does not elaborate on exactly what this was, he appears to referring to the crêpe ring he had invoked before). Only further study, in his opinion, would reveal which of the four explanations was the correct one.

The final contribution from the *Astro-Physical Station, South Birmingham* during this period was a description of a mysterious faint luminous beam, some 80° long and ½° wide, traversing the sky horizontally in the south on the evening of 1905 August 18. Its identity remains a mystery. He had ruled out a searchlight or a cirrus cloud and noted is similarity to the tale of a comet (103).

### *Light*: the great driving force of the universe

Very few contributions from Packer were published in the *English Mechanic* after about 1907, whereas until that time, he was a prolific contributor. One wonders whether he had become disenchanted by the way his ideas had been received by the astronomical world, especially the criticism he received as a result of his work on the solar corona. He also made no further contributions to the BAA *Journal*, which as we have seen had declined to publish some of his submissions. In 1907 concern was expressed by some members that the BAA, and its *Journal,* has lost its way since it was established in 1890 and membership was beginning to slide. Whilst others wrote in support of the BAA, Packer revealed his own view on the matter: "Everyone who has the interest of astronomy at heart, and the progress of that representative of amateur astronomy, the British Astronomical Association, in particular, must regret its present decayed state" (104).

Packer's last contrition to the *English Mechanic* was published in 1923 September and described a bright state of Betelgeuse, one of his favourite subjects (105). He continued to maintain an interest in astronomy, although this was more on the theoretical aspects of the subject. He published three *Manifestos*, as he called them, on the nature of light and the Universe in 1931 (106), 1932 (107) and 1935 (108). Central to his model of the universe was that its structure was maintained by two opposing forces, gravity and "light – the great driving force of the universe". Reflecting on the observed stability of the positions of stars relative to each other, "it became increasingly obvious to me that we were in the presence of some mighty cosmic force as great as but more far reaching than gravitation and of entirely opposite character. For whereas the virtue of gravitation is attraction, the virtue of this new force [that is, *light*] must be repulsion or propulsion" (107) (the emphases are Packer's).

He went on to describe a thought experiment: if light were a repulsive or propulsive force, could its effects been seen and measured? His first piece of evidence for a "*light drive*" was the fact that comet tails point away from the sun. If the sun, a relatively small star in the scheme of things, could exert such an effect, what would be the effect of many larger stars on the solar system? Since the brightest stars would have the largest effect, by virtue of their greater light output, with Sirius and Canopus having the greatest effect of all, he used simple arithmetic and vectors to calculate the combined effect of the 88 brightest stars on the sun (Figure 10).This showed that the light pressure was not evenly balanced and, following Newton's laws, if forces were not balanced the object upon which they act should move. Based on this assumption, he calculated the direction of movement that could be expected from the nett force. Here was the proof: the direction of movement was within three degrees of the position of the solar apex, the apparent direction in which the sun is moving, that is towards a point in Hercules.

Although Packer's idea of light propulsion might have been off the mark, it is known that light, and other electromagnetic radiation, does exert pressure when photons impinge on a surface (109). Whilst it is not strong enough for neighbouring stars to affect the course of the solar system through space, it is nevertheless a factor that is taken into account when planning the trajectory of interplanetary space probes the course of which can be perturbed by solar radiation. Solar radiation pressure has even been suggested as a source of propulsion for space vehicles equipped with sails, which is perhaps not too distant from Packer's concept of a "light drive" (110). Radiation pressure is also important in dispersing molecular clouds around newly formed stars in interstellar space and in forming the dust tails of comets.

Packer sent copies of his *Manifestos* to libraries around the world, including the University of Cambridge, the British Museum, the Royal Society of Edinburgh, the Physical Society (London), the Mount Wilson Observatory and, of course, the Birmingham Central Library (111). Less than a year after completing his third Manifesto, he passed away at Rednal in 1936 March.

**Perspectives**

With the benefit of hindsight, it can be seen that some of Packer's ideas were naive and some of his hypotheses were strange to say the least. Moreover the validity of some of his observational work, such as his meteor reports, was questionable, and some, such as his photographic work on the solar corona, was fundamentally flawed. There is also a perception that he was keen for recognition by the astronomical community, which might have prompted him to propose unifying theories, such as his *Cosmical Law* on the link between meteors and novae (112), and his theories about the existence of a *light drive*. He was certainly disappointed when his ideas were not taken up by others, but he still held onto his beliefs even in the face of contradictory evidence from the astronomical community. Thus, his third *Manifesto*, contains an addendum on his experiments conducted in 1935 January showing that metal-penetrating rays from the sun can be captured on photographic paper, although he did not go so far as to claim the resulting images definitely represented the solar corona.

However, we should be very careful not to dismiss the entirety of Packer's work because of these apparent shortcomings. There can be no doubt that he was an enthusiastic observer who achieved much success with the limited means at his disposal. As we have seen, not only did he discover new variable stars, including only the second periodic variable to be found in a globular cluster, but he might also have witnessed an important event in M77.

What sets him apart from many of his contemporaries, who like him observed the night sky in their spare time, was his desire to take observational data and develop hypotheses from them, such as his proposed links between comets and the solar cycle, and between meteors and novae and between meteors and earthquakes. Moving along the path from observation, to analysis, to hypothesis, to prediction is how science progresses. Thus Packer had the intellectual curiosity and the strength of character to put forward a hypothesis, to set out the evidence on which it is based and to allow others come forward with data that support or refute it.

Largely self-taught, Packer was evidently stimulated by the latest developments in the physical sciences that were rapidly unfolding at the time and which he would have learnt about from the journals and books he had access to through his profession as librarian. These included breakthroughs in the new science of astrophysics, based on advancements in spectroscopy and photography, alongside new insights into the nature of light and the electromagnetic spectrum, including the discovery of X-rays. He was not content merely to read about these developments, noting them from a distance. Instead in some cases they became the tools with which he worked, even though his approach was misguided on occasions, often because he lacked a deep understanding of the underlying science.

Packer's characteristic determination can be seen reflected in a life that took him from the relatively humble origins of a family of basket makers, via employment as oilman's assistant on the streets of London, through professional qualifications and on to the respected role of librarian in Birmingham. And yet at the same time he had the perseverance to pursue an interest in astronomy to a level which brought him to the attention of some of the most famous astronomers of the age, including E.E. Barnard and E.C. Pickering. One wonders what else he might have achieved had he had greater means at his disposal and with somebody to guide his research, perhaps as a member of a research institution rather than working on his own.

It is perhaps fitting to conclude with Packer's own words which sum up his motivation for astronomy: "the true enthusiast…..will find in the very act of investigating and exploring the heavens an endless and unspeakable delight and pleasure, no matter how humble his means, apart from the lasting satisfaction which arises from the discovery of some new truth as a resultant, all of which infinitely outweighs any outside reward, however ample" (113).

**Acknowledgements**

Many people have provided an enormous amount of assistance with this research. Sheila Smail, D.E. Packer's great granddaughter, of Ottawa, Canada, provided important information about her family's history from her own research on the topic. She put me in touch with Professor Barbara Becker, University of California at Irvine, who shared William Huggins's correspondence concerning Packer's work on the solar corona and which will appear in her forthcoming book, *Selected Correspondence of William Huggins* (London: Pickering & Chatto Publishers). Both Sheila and Barbara provided much encouragement to pursue this research. Jean Poyner kindly researched Packer's papers in the Library of Birmingham. John Bortle, Gary Poyner, Bob Argyll. John Toone and Mike Frost were helpful on matters regarding M77. Tony Markham gave advice about nebulous meteors. Leoš Ondra and Professor Gáspár Bakos, Princeton University, gave permission to reproduce their figures relating to the variables in M5. I am most grateful to all these people for their help and kindness.

During my research I made extensive use of scanned back numbers of the BAA *Journal*, which exist largely thanks to the herculean scanning efforts of Sheridan Williams, as well as of the *English Mechanic*, thanks to Eric Hutton. These are truly wonderful resources for historians of nineteenth and twentieth century British astronomy. I also used the NASA/Smithsonian Astrophysics Data System, the AAVSO Variable Star Index, and SIMBAD, operated through the Centre de Données Astronomiques (Strasbourg, France).

**Notes and references**

1. Birthplace: Mercy Cottage, New St., South Bermondsey, London.

2. This was Edward Packer's second marriage. His first wife, Mary Ann Packer formerly Jordan, died in 1859. D.E. Packer had three younger sisters: Emma Bidmead Packer (b. 1863), Fanny Bidmead Packer (b. 1866) and Ruth Packer (b. 1874).

3. His school report for the term ending 1873 July shows that he came third in a class of 23 boys in arithmetic, algebra and general mathematics (rated "very good"). He attained the same ranking in writing ("excellent"). Latin was the only subject in which apparently did not thrive, coming seventeenth.

4. Smail S., Personal communication (2014). Sheila Smail, David Elijah Packer's great granddaughter, has researched the family history using original sources including Census entries.

5. 1881 Census. At this time he was still living with his parents at 22 Queen's Head, Horsleydown, London.

6. 1891 Census. Address: 217 Tooley Street, Horsleydown, London.

7. Packer wrote at least one letter to the English Mechanic from Cambridge. This was on observations of the aurorae and zodiacal light in 1892 January to March. Packer D.E., English Mechanic, 1408, 81 (1892).

8. Smith L., "Biographical notes for David Elijah Packer" (1995). Unpublished manuscript containing biographies of D.E. Packer and his wife, including family reminiscences. It was written by Packer's granddaughter, Lynda Smith. Lynda was the daughter of Ruth Ida Smith (nee Packer) and much of the information in the manuscript is attributed to Ruth.

9. Parish of St George the Martyr, Southwark.

10. Roland Packer (1895-1974), Howard Packer (1896-1977), David Edward Packer (1898-1984), Sarah J. (Packer) Taylor (1898-1988), Emma M. Packer (b. 1902), Bernard Packer (1904-1969) and Ruth I. (Packer) Smith (1906-1990).

11. Packer D.E., JBAA, 4, 96 (1894).

12. He was at the Central Library from sometime in late 1893 until at least 1911, when his Census return places him there. The 1921 Census has him at the Stirchley & Bournville Free Public Library. Moreover, in 1921 the UK, Midlands and Various UK Trade Directories, 1770-1941, lists his profession as "Librarian at the Stirchley & Bournville Free Public Library". Family recollections have him serving at Kings Norton, Northfield and Harborne.

13. Packer D.E., English Mechanic, 1212, 351 (1888). This was Packer's first letter to the English Mechanic, the edition being published on 1888 June 15.

14. Ingall owned several dialyte refractors, the largest being 10-inch (25.4cm). He used a 4½-inch dialyte to observe the lunar crater Plato in 1865 (Ingall H., AReg, 3, 140-141 (1865)). One wonders if this was the same telescope that Packer later used. Packer makes several references to his visits to Ingall's house to observe with Ingall's 10-inch dialyte.

15. Ingall's cat was the subject of some discussion in the English Mechanic for its longevity (19 years 7 months) and fecundity (she had more than 100 kittens) as described by Ingall's son who inherited the feline upon his father's death; Ingall M.A., English Mechanic, 2055, 19 (1906). "She used to sit on the high steps of his large telescope, and while he observed, she observed too, and sat up with him till the small hours of the morning".

16. In a letter to George Ellery Hale of 1896 March 10, Huggins writes about Packer: "You may remember him as the doorkeeper at the rooms of the Royal Astronomical Society". Packer was clearly knowledgeable about the documents kept in the RAS Library as he helped W.T. Lynn locate a manuscript there "in the absence of Mr. Wesley", W.H. Wesley (1841-1922) being the Assistant Secretary of the RAS for many years. See: Lynn W.T., Obs., 14, 345-346 (1891).

17. Packer D.E., English Mechanic, 2063, 206 (1904). He wrote several letters on how to popularise astronomy and suggested setting up evening classes for workers in Britain's major cities.

18. Common A.A., MNRAS, 50, 517 (1890).

19. Packer D.E., English Mechanic, 1318, 378 (1890).

20. Packer D.E., SidM, 9, 381 (1890). On checking his notes, Packer found that he had observed the star as far back at 1889 May 31.

21. Packer D.E., English Mechanic, 1322, 462 (1890).

22. Pickering E.C., AN, 125,157 (1890). As was standard practice at Harvard, the announcement was communicated by Pickering, as Director, even though the work was done by Fleming. The Harvard announcement was also carried in the Sidereal Messenger: Fleming W., SidM, 9, 380 (1890).

23. Packer D.E., English Mechanic, 1305, 94-95 (1890).

24. Packer D.E., English Mechanic, 1330, 80-81 (1890). The pseudonymous "Another FRAS" suggested that Packer should have reported his observations to Nature. Whilst Packer clearly had priority for the discovery he was evidently irked by the discussion. Packer testily observed: "This tendency in certain quarters to snub the work of amateurs is heartily to be deprecated. Let us hope the new British Astronomical Association may crush this tendency, and give each persevering amateur the justice which is his due, and so encourage him to prosecute his researches with refreshed energy". Packer D.E., English Mechanic, 1338, 248 (1890).

25. Barnard E.E., AN, 147, 243-248 (1898). In spite of Barnard's endorsement, there still remained some doubt about Packer's second variable (84 in the Harvard list) as Harvard's Solon Bailey (1854-1931) suspected that Packer might have observed a blend of three stars. See: Bailey S.I., Annals of Harvard College Observatory, 78, 99-194 (1917).

26. Bailey S.I., op. cit.

96. The fate of these plates is not known. Enquiries amongst Packer's descendents and the Library of Birmingham have drawn a blank. Naturally, if anyone has information about them, the author would be delighted to hear from them.

97. Packer D.E., English Mechanic, 2007, 114-115 (1903). In this letter he laments "All this time I was working in the dark, for the Röntgen rays had not yet been discovered, nor their existence even dreamt of, and I found it extremely difficult to get professional men to believe that there existed rays capable of penetrating substances opaque to ordinary light". However, X-rays were already announced and Huggins referred to X-rays in his comments on Packer's work in 1896 January.

98. Hollis H.P., English Mechanic, 2838, 42 (1919).

99. Packer D.E., English Mechanic, 2102, 497-498 (1905).

100. Packer D.E., English Mechanic, 2103, 520 (1905).

101. Packer D.E., English Mechanic, 2104, 544 (1905).

102. Packer D.E., English Mechanic, 2106, 592 (1905).

103. Packer D.E., English Mechanic, 2110, 88 (1905). There was a further contribution from the Astro-Physical Station, South Birmingham in 1906 February, but it was not on the theme of nebulous clouds. In the letter Packer advocated a more thorough study of electrical storms and wondered if there might be a connexion to the aurorae: Packer D.E., English Mechanic, 2136, 87 (1906).

104. Packer D.E., English Mechanic, 2230, 450-451 (1907).

105. Packer D.E., English Mechanic, 3050, 85 (1923).

106. Packer D.E. The Stellarium: a miniature model representation of the sidereal universe, and the wonders it reveals, 1931. Typescript kept in the Archives of the Library of Birmingham.

107. Packer D.E. First fruits of Stellarium research. Light: the great driving force of the universe, 1932. Typescript kept in the Archives of Birmingham Central Library.

108. Packer D.E. Radiant force: the true light-force that moves the worlds, 1935. Typescript kept in the Archives of Birmingham Central Library.

109. James Clerk Maxwell (1831-1871) described the pressure that electromagnetic radiation would have on surfaces exposed to it in 1862. It was proven experimentally by Russian physicist Pyotr Lebedev (1866-1912) in 1900, and by Ernest Fox Nichols (1869-1924) & Gordon Ferrie Hull (1870-1956) in 1901.

110. The concept was mentioned by Jules Verne in 'From the Earth to the Moon'. In 2010, the Japan Aerospace Exploration Agency's IKAROS project successfully deployed a solar sail which succeeded in propelling its payload.

111. Current members of the Packer family have copies of letters from these institutions acknowledging receipt of the third Manifesto.

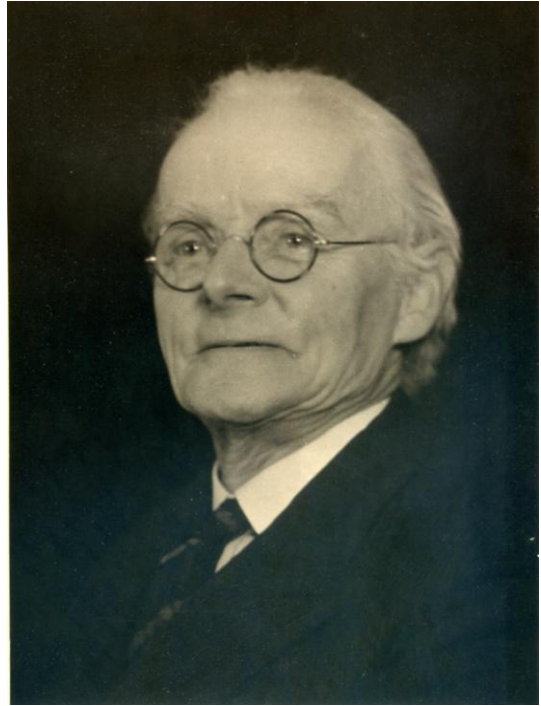

Figure 1: David Elijah Packer in later life

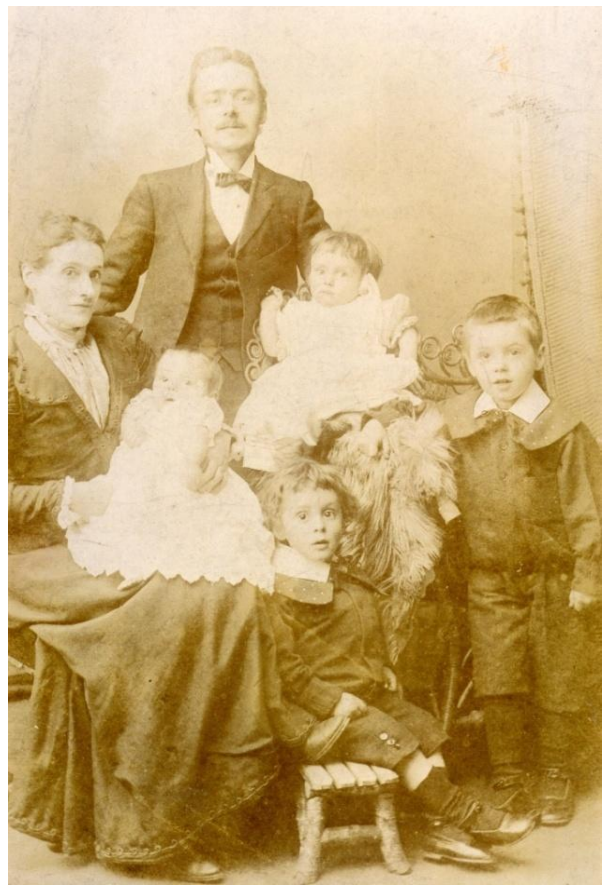

Figure 2: D.E. Packer and his family in 1899. Daughter, Sarah, is in the arms of Packer's wife, Sarah Ann. David Edward is next to his father. Howard sits on the stool and Roland is standing on the right.

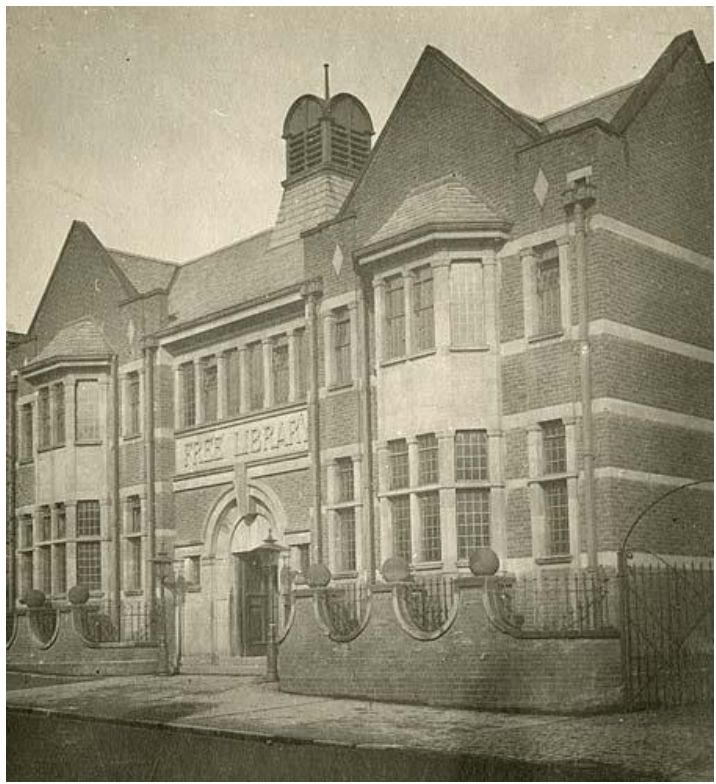

Figure 3: Stirchley & Bournville Free Public Library, 1913 (114). (Used with permission of the Library of Birmingham Photographic Collections. Photographer: Lewis Lloyd)

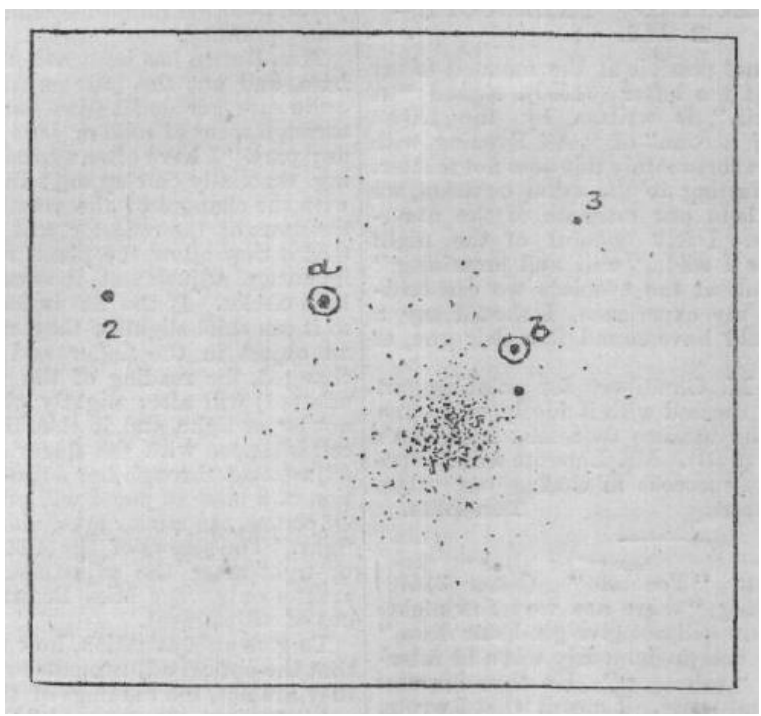

Figure 4: Packer's sketch of his two new variables in M5. From reference (115). North down, west to left.

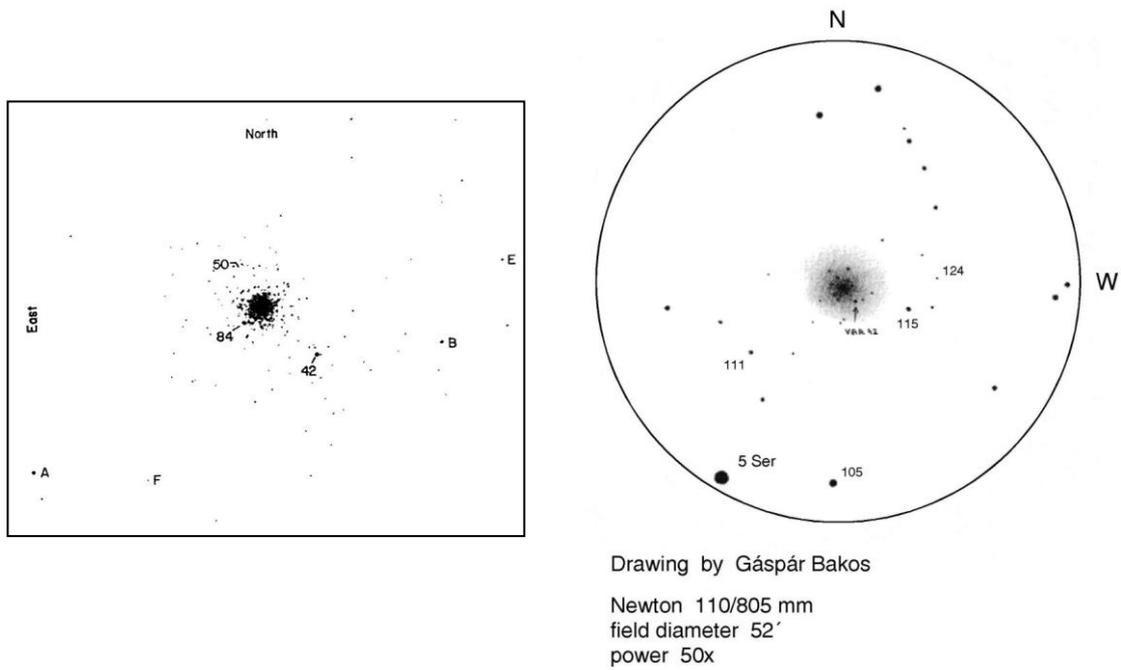

Figure 5: Variable stars in M5 (a) – left – Photograph from reference (31); (b) - right- drawing by Gáspár Bakos

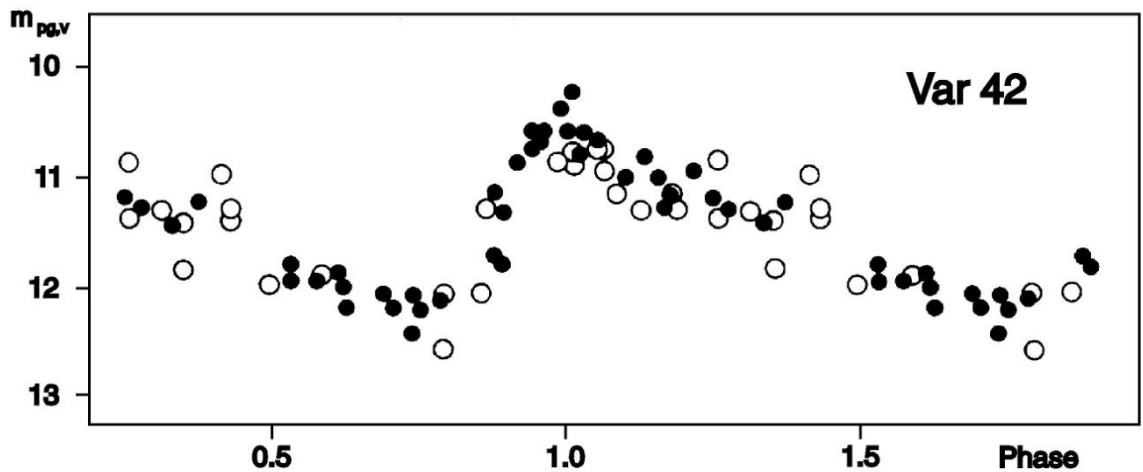

Figure 6: Phase diagram of M5 Variable 42. Filled circles represent photographic data given in reference (31);open circles are visual estimates by Jirka Dusek, Kamil Hornoch and Leoš Ondra. The data folded on a period of 25.738 days (plot by Leoš Ondra)

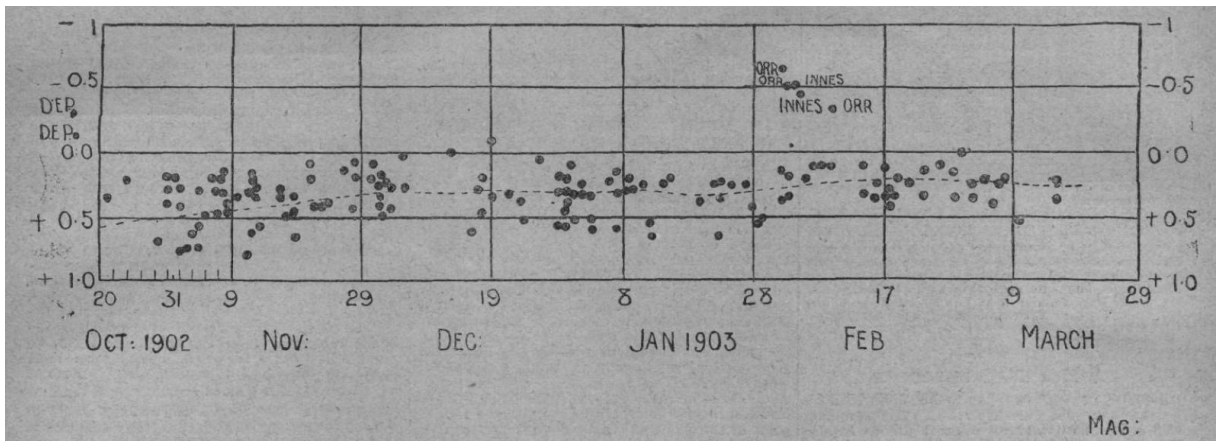

Figure 7: Light curve of Betelgeuse between 1902 October and 1903 March drawn by E.E. Markwick. Observations by Packer ("DEP"), R.T. Innes and M.A. Orr are annotated. The other observations were from VSS observers (Messrs. Besley, Brooks, Child, Le Beau, Oakes and Markwick). From reference (116)

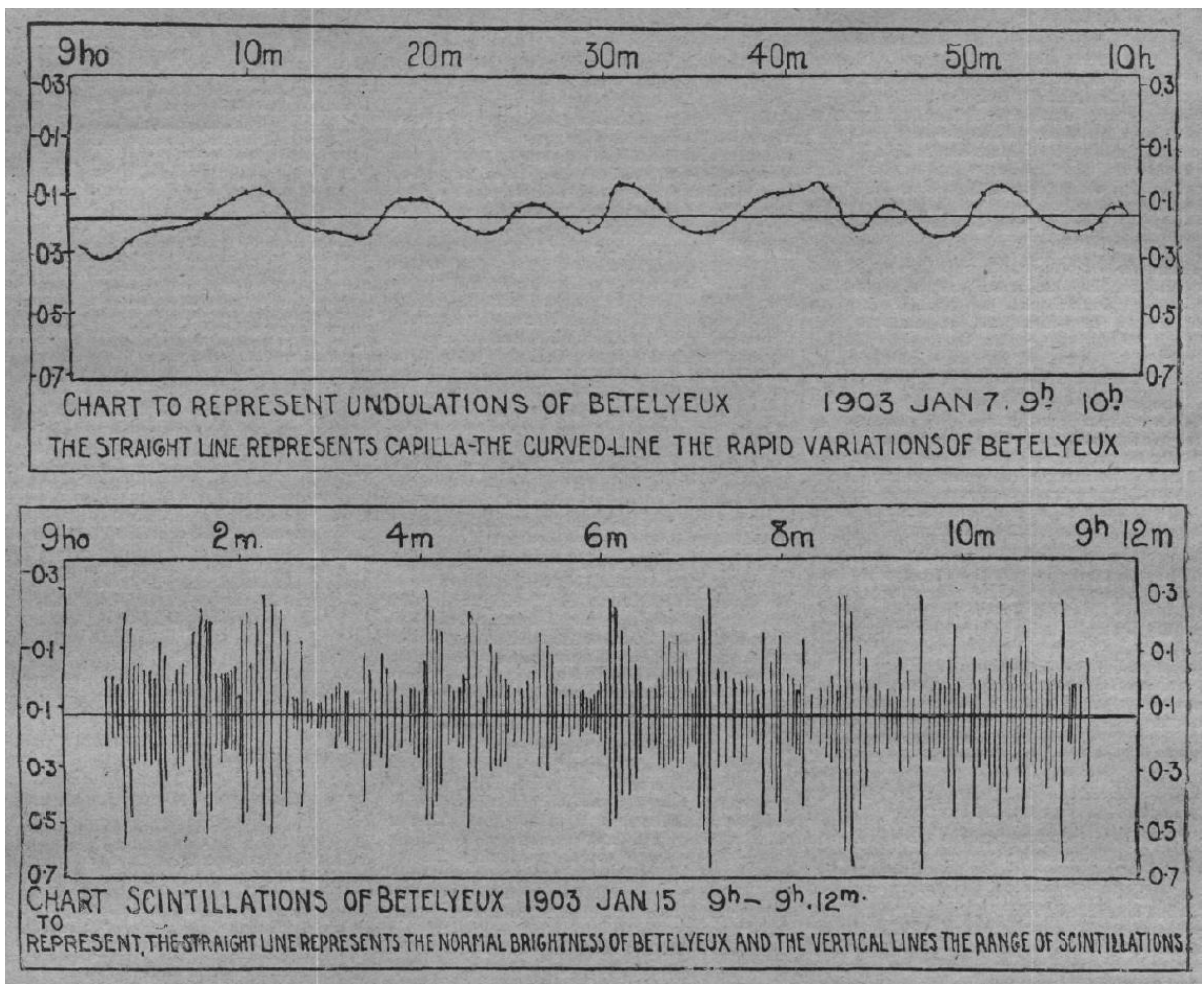

Figure 8: Scintillation of Betelgeuse 1903 Jan 7 and Jan 15. From reference (42)

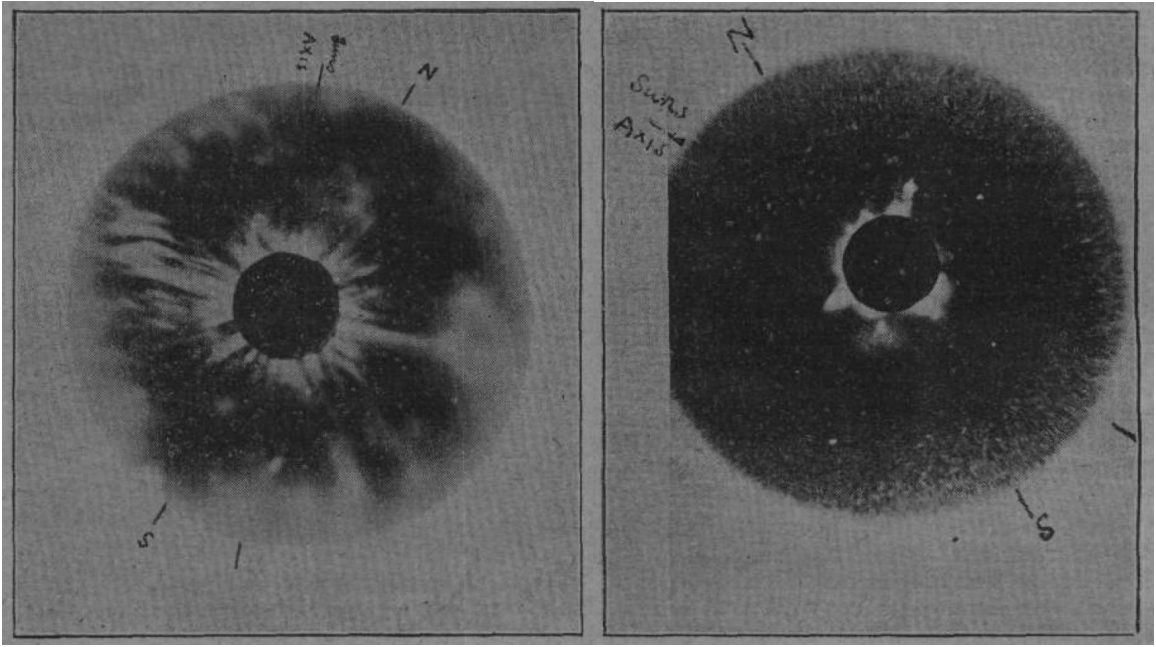

Figure 9: Packer's photographs of the solar corona

Left: 1895 August 27 (exposure 60 seconds), Right: 1895 September 25 (exposure 4 minutes). "The two photographs submitted were taken in a small camera with a clear aperture (pin-hole) of $^1/_{20}$-in. diameter, and a depth of 5in., upon ordinary lantern-plates, covered with lead-foil". From reference (85).

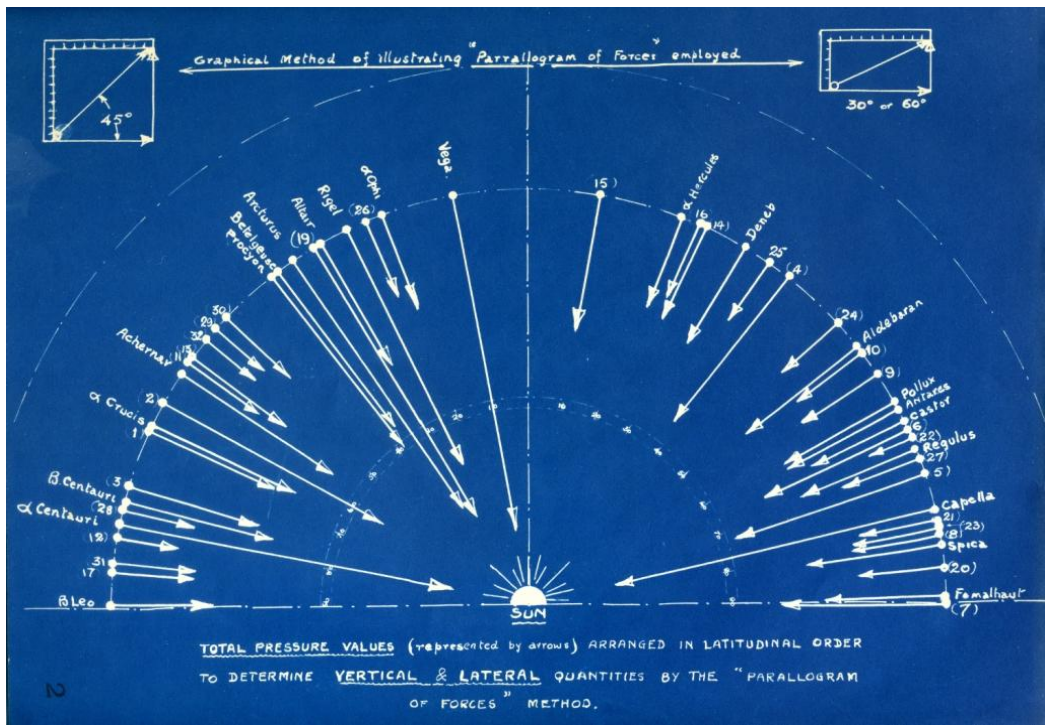

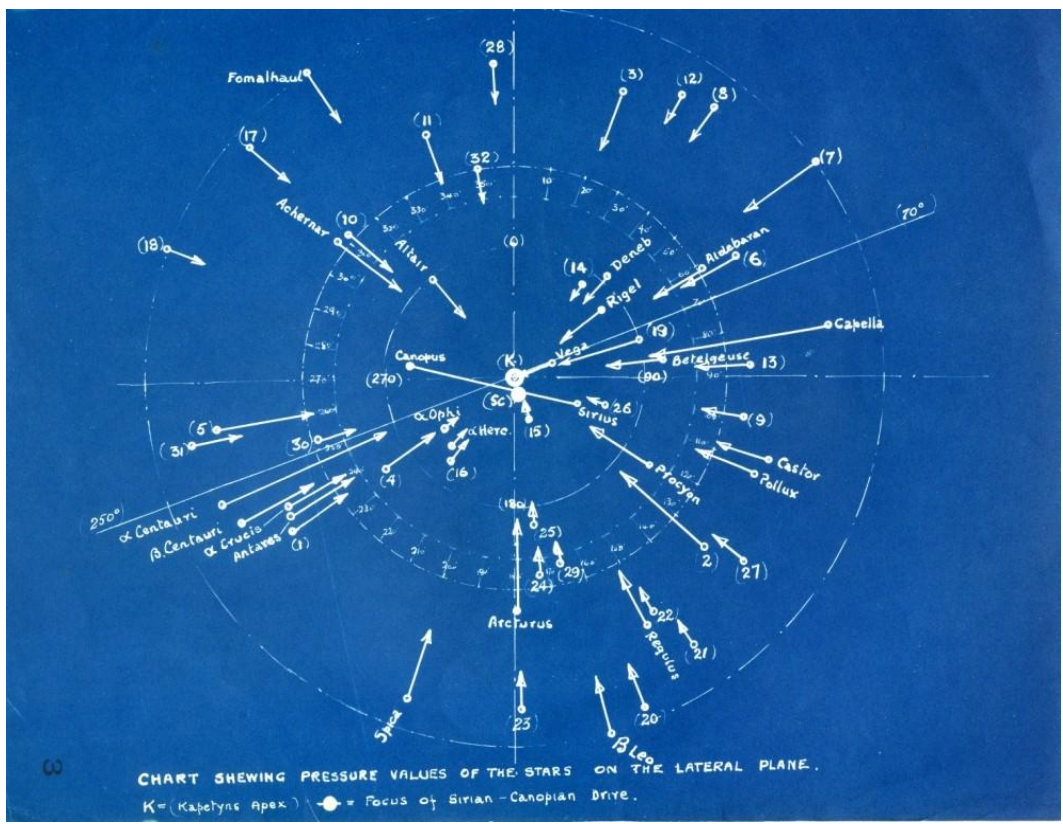

Figure 10: Packer's diagrams showing the effect of light pressure exerted by the brightest stars on the sun. The length of the arrow represent the magnitude of the pressure. Top: Vectors in the vertical plane. Bottom: Vectors in the horizontal plane. From reference: (107)